\documentclass{article}
\usepackage{gensymb}
\usepackage{amssymb,euscript,latexsym,amsmath}
\usepackage{graphicx}
\usepackage[dvips]{epsfig}
\usepackage{color}
\usepackage{epstopdf}
\usepackage{setspace}
\usepackage{subfigure}
\usepackage{lipsum}
\usepackage{natbib}
\usepackage{ulem}
\usepackage[margin=1.2in]{geometry}

\usepackage{caption2}

\DeclareGraphicsExtensions{.eps,.ps,.pdf}

\newcommand{\conj}[1]{\bar{#1}}

\begin{document}
\normalem


\title{Fishlike Rheotaxis} 
\author{Brendan Colvert and Eva Kanso} 

\maketitle

\begin{abstract}
Fish rheotaxis, or alignment into flow currents, results from intertwined sensory, neural and actuation mechanisms, all coupled with hydrodynamics to produce a behavior that is critical for upstream migration and position holding in oncoming flows. Among several sensory modalities, the lateral line sensory system is thought to play a major role in the fish  ability to sense minute water motions in their vicinity and, thus, in their rheotactic behavior. Here, we propose a theoretical model consisting of a fishlike body equipped with lateral pressure sensors in oncoming uniform flows. We compute the optimal sensor locations that maximize the sensory output. Our results confirm recent experimental findings that correlate the layout of the lateral line sensors with the distribution of hydrodynamic information at the fish surface. We then examine the behavioral response of the fishlike model as a function of its orientation and swimming speed relative to the background flow. Our working hypothesis is that fish respond to sensory information by adjusting their orientation according to the perceived difference in pressure. We find that, as in fish rheotaxis, the fishlike body responds by aligning into the oncoming flow.  These findings may have significant implications on understanding the interplay between the sensory output and fish behavior.

\end{abstract}

\section{Introduction}

Fish exhibit remarkable abilities to sense and respond to hydrodynamic signals. A chief example is fish rheotaxis where fish actively sense water motions and respond either by aligning their bodies in the direction opposite to the oncoming flow, such as during upsteam migration and position holding, or by going with the flow to flee high-flow regions; see, for example,~\cite{Arnold1974}.
In addition to alignment with or against the flow, the fish capability to discern the information contained in various flow patterns is manifested in behaviors such as obstacle avoidance~\cite{Windsor2010}, energy extraction from ambient vortices~\cite{Liao2003}, and identification and tracking of flow disturbances left by prey~\cite{Pohlmann2001}. These behaviors have been reported even in the absence of visual cues, as in the extreme case of the  blind Mexican  cave fish~\cite{Montgomery2001, Windsor2010}.

\begin{figure}
\centering
\includegraphics[width=6in]{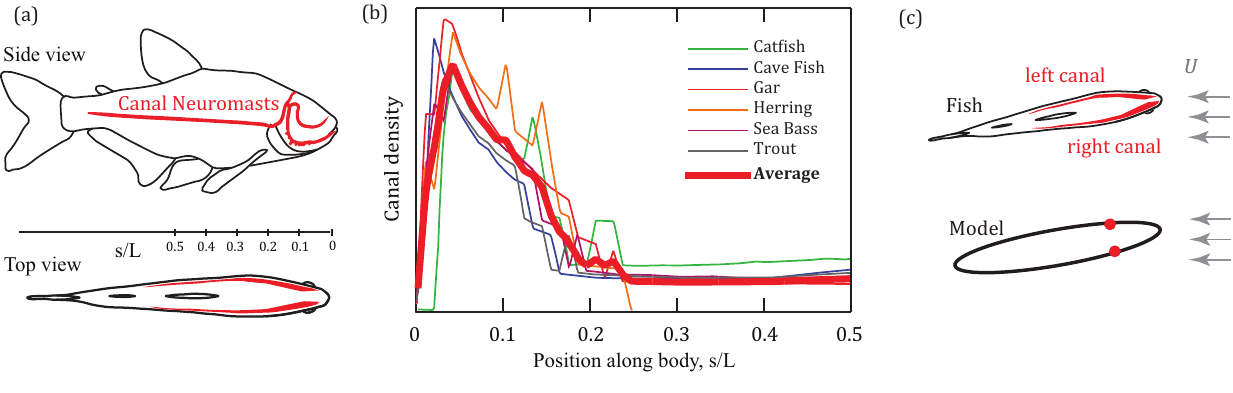}
\caption{(a) Spatial distribution of the  lateral-line canal neuromasts  in {\em Astyanax fasciatus} (Blind Mexican cave fish), inspired from \cite{Montgomery2001}. (b) Canal density in various species of fish, adapted from \cite{Ristroph2015}. (c)  Lateral-line canal seen from the top and corresponding model system proposed here.} 
\label{fig:lateral-lines-schematic}
\end{figure}

The fish ability  to sense minute water motions in their vicinity is attributed to the  lateral-line mechanosensory system ~\cite{Engelmann2000}. The lateral line system consists primarily of two types of sensors: superficial and canal neuromasts. Superficial neuromasts are located in the skin and are functionally suited for measuring velocity~\cite{Kroese1992}. Canal neuromasts are recessed in bone-like canals and are thought to measure pressure~\cite{Coombs2005}. 
 The layout of the canal sensory system is highly conserved across diverse species and evolutionary time~\cite{Coombs1988}; namely,  it extends over much of the body but is concentrated near the head~\cite{Siregar1994}; see figure~\ref{fig:lateral-lines-schematic}.
 This basic arrangement also holds in the blind Mexican  cavefish, albeit with more  sensory receptors in comparison to their surface-dwelling relatives who retained a functional visual system~\cite{Yoshizawa2013}.

 The universality of the lateral line architecture suggests a functional advantage in having higher density  of sensors in the anterior segment of the fish. 
 Recent experiments by Ristroph, Liao and Zhang correlate anatomical measurements of canal density with pressure measurements from flow experiments~\cite{Ristroph2015}. They suggest that the relevant sensory output is the difference in pressure between the left and right canals, rather than the absolute pressure itself. 
 They further show that the canal system is concentrated at locations on the body that experience strong variations in this pressure difference. While these findings provide valuable insights into the relevant quantity that is sensed, a deeper understanding of how behavior emerges from sensory information requires, in  addition to identifying the sensory output, a knowledge about how such sensory information is processed and translated into decision making and behavior.  Discerning how sensory signals translate into behavior in live fish is a complicated task.  Behavior is the result of intertwined complex mechanisms, where sensory feedback and neural control, muscular actuation and preflexes, and hydrodynamics are combined to produce the remarkable ability of fish to respond to hydrodynamic cues.

In this work, we propose a mathematically-tractable model system  that allows us to shed light on some of these questions. Namely, we investigate the optimal sensor placement that maximizes the sensory output, that is, the difference in pressure between the left and right sensors, and how behavior emerges in light of this sensory information. Our model consists of a planar elliptic body swimming at a constant speed along its major axis, and equipped with pressure sensors located laterally on the two sides of the ellipse in a symmetric configuration about its major axis, see figure~\ref{fig:lateral-lines-schematic}c. The sensors provide one sensory output, namely, the difference in pressure between the body's two sides, which is consistent with experimental findings of~\cite{Ristroph2015}. We postulate that the body responds to the sensory output by adjusting its orientation. Our goal is to test whether this hypothesis or `control law' is sufficient to capture the rheotactic behavior exhibited by many fish species. To this end, we place the fish model in oncoming uniform flows in an otherwise unbounded domain of inviscid and incompressible fluid. We investigate optimal sensor placement in these flows and examine the response of the body to pressure cues. Interestingly, the body responds by aligning with or against the flow as observed in fish rheotaxis. This rheotactic behavior depends crucially on the orientation of the fishlike body and its swimming speed relative to the oncoming flow.  We conclude by commenting on the relevance of these findings to understanding both the sensory response in fish and the conceptual design of engineered underwater sensory systems.

\section{Model}

\begin{figure}
\centering
\includegraphics[width = 6in]{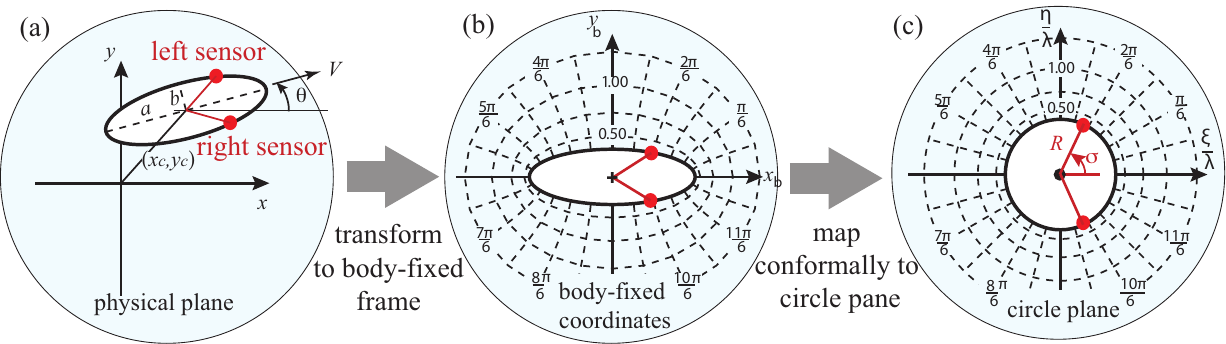}
\caption{ Schematic of the model system  (a) in the physical plane, (b)  in a body-fixed frame, with dashed lines corresponding to level sets of constant elliptic coordinates; and (c) conformally mapped to  the circle plane, with dashed lines corresponding to level sets of constant polar coordinates.
\label{fig:schematic}}
\end{figure}

Consider a planar elliptic body with semi-major axis $a$ and semi-minor axis $b$, submerged in an otherwise unbounded domain of inviscid and incompressible fluid.  Let $(x_c , y_c)$ denote the position of the center of the body with respect to a fixed inertial frame with Cartesian coordinates $(x,y)$. Let $\theta$ denote the orientation angle measured from the positive $x$-direction to the ellipse's major axis as shown in figure~\ref{fig:schematic}a.  The linear and angular velocities of the body are given by $(\dot{x}_c , \dot{y}_c)$ and $\dot{\theta}$, respectively. Here, the dot $\dot{()}$ corresponds to derivative with respect to time $t$. 
To emulate the fish lateral line, we equip the body with two pressure sensors located at  $(x_l,y_l)$ and $(x_r,y_r)$ in a symmetric configuration about the major axis of the ellipse. These sensors capture the local value of the pressure field $p(x,y)$ to provide one sensory output, namely, the difference in pressure $\Delta p =  p(x_l,y_l) - p(x_r,y_r)$.

We investigate the dependence of the sensory output $\Delta p$ on body orientation $\theta$ and swimming velocity $V$  in uniform flows of speed $U$ in the negative $x$-direction. Analytic expressions for the fluid velocity and pressure fields are obtained by conformally mapping the infinite fluid domain outside the elliptic body into  the domain outside a circle of radius $R = (a+b)/2$. For convenience, we introduce the complex coordinate $z = x+ {\rm i} y$, where ${\rm i} =  \sqrt{-1}$, in the physical plane, such that the location of the center of the body is given by $z_c = x_c + {\rm i} y_c$ and the locations of the sensors are $z_r = x_r + {\rm i} y_r$ and $z_l = x_l + {\rm i} y_l$. These coordinates are then transformed to a body-fixed frame (figure~\ref{fig:schematic}b) prior to mapping the ellipse plane conformally to the circle plane (figure~\ref{fig:schematic}c). The conformal mapping from the circle to the physical plane takes the form (see, e.g.,~\cite{Michelin2009,Ysasi2011}),
\begin{equation}\label{eq:conform}
{z}   = z_c + \left(\zeta+\frac{\lambda^{2}}{\zeta} \right)e^{i\theta}.
\end{equation}
Here,  $\zeta = \xi + i \eta$ is the complex variable in the circle plane and $\lambda = (\sqrt{a^2-b^2})/2$ is a measure of the ellipse's eccentricity, whereas $e^{(\cdot)}$ denotes the exponential function.


\begin{figure}
\centering
\includegraphics[width=6in]{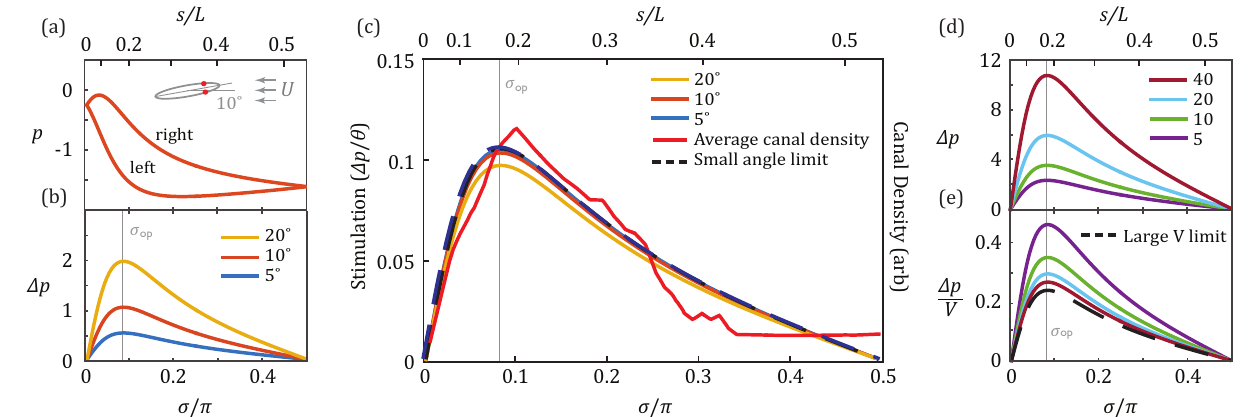}
\caption{(a) Pressure distributions along the body for $b/a = 0.266$, $\theta = 5\degree$, $V=0$. (b) Pressure difference $\Delta p$ in~\eqref{eq:dp} as a function of $\sigma$ for $\theta = 5\degree, 10\degree,$ and $20\degree$ and $V=0$. (c) Stimulation ($\Delta p/\theta$) for $\theta = 5\degree, 10\degree,$ and $20\degree$ and its theoretical limit obtained by using small $\theta$ in~\eqref{eq:dp} compared with the average canal density of figure~\ref{fig:lateral-lines-schematic}(b). (d) Pressure difference as as a function of $\sigma$ for $V=5, 10, 20, 40$ and $\theta = 5\degree$. (e) $\Delta p/ V$ for same parameter values.} 
\label{fig:dp}
\end{figure}

Let $u(z)= u_x(x,y) + {\rm i} u_y(x,y)$ denote the complex fluid velocity in the physical plane. The corresponding complex potential function can be written as $F(z) = \Phi(x,y) + {\rm i} \Psi(x,y)$, with $\Phi$ being the real potential function and $\Psi$ the stream function. The existence of $\Phi$ and $\Psi$ is guaranteed by the irrotationality and incompressibility conditions;
see, e.g.,~\cite{Lamb1932}. By linearity of the problem, the complex potential $F(z)$ can be written as a superposition of two contributions, $F  =F_{\rm b} + F_{\rm u}$, where $F_{\rm b} (z) =  -V{(R^2-\lambda^2)}/{\zeta}$ is due to the motion of the elliptic body and $F_{\rm u}(z) =  -U e^{i\theta} \zeta - Ue^{-i\theta}  {R^2}/{\zeta}$ is due to the external uniform flow $U$.
Note that we neglected the effect of the turning rate $\dot{\theta}$ on $F_{\rm b}$, considering only quasi-static changes in body orientation to best emulate the experimental procedure of~\cite{Ristroph2015}. 
We introduce non-dimensional counterparts to the above quantities using the characteristic length scale $\lambda$ and time scale $\lambda/U$. The non-dimensional radius  $R$  of the circle is given by $R = (a+b)/(2\lambda) = \sqrt{{(a+b)}/{(a-b)}}$, and all parameters and variables are now considered  in non-dimensional form. To this end, the dimensionless stream function takes the form
\begin{equation} \label{eq:uniform}
F =  F_{\rm b} + F_{\rm u} = V \frac{1-R^2}{\zeta}-\zeta e^{i\theta}-\frac{R^2}{\zeta}e^{-i\theta} .
\end{equation}
The complex conjugates $\conj{u}(\zeta) = u_\xi - i u_\eta$ and $\conj{u}(z)=u_x - i u_y$ of the fluid velocity field in the circle and physical planes are given by 
\begin{equation}\label{eq:vel}
\conj{u}(\zeta)= \dfrac{dF(\zeta)}{d\zeta},  \quad \conj{u}(z)= \dfrac{dF}{d\zeta}\dfrac{d\zeta}{dz}.
\end{equation}
The pressure field $p$ is obtained from the steady Bernoulli equation  $p - p_{\infty}=-\frac{1}{2}|u|^2,$
where $p_\infty$ is the pressure at infinity and is undetermined.
The sensory output is given by the difference in pressure between the left and right sensors 
$\zeta_{l}$ and $\zeta_{r}$,
\begin{equation}\label{eq:dp0}
\Delta p = p (\zeta_l)-p(\zeta_r) = \frac{1}{2}|u|_{\zeta_r}^2 - \frac{1}{2}|u|_{\zeta_l}^2.
\end{equation}
The location of the right/left sensors in the circle plane is expressed as $\zeta_{r} = R e^{{\rm - i}\sigma}$ and $\zeta_{l} = R e^{{\rm  i}\sigma}$, with $\sigma$ being the sensors angular position (figure~\ref{fig:schematic}c). 
Alternatively, one could use the arc-length $s$ normalized by the length $L$, as done in~\cite{Ristroph2015} (see figure~\ref{fig:lateral-lines-schematic}a). In the context of the elliptic model, one can put the normalized arc-length $s/L$ in one-to-one correspondence with $\sigma$, as noted in figure~\ref{fig:dp}. Hereafter, we use $\sigma$ to refer to the location of the sensors.

We assume that the fishlike body swims at a constant speed $V$ in the direction of its major axis of symmetry (figure~\ref{fig:schematic}a), and responds to  sensory information by adjusting its orientation depending on the value of the pressure difference $\Delta p$.  Namely, we describe the body response to sensory output using the following kinematic model
\begin{equation}
\begin{split}
\dot{x}_c  = V\cos \theta, \quad 
\dot{y}_c  = V\sin \theta, \quad
\dot{\theta} = \kappa\Delta p .
\label{eq:eom}
\end{split}
\end{equation}
Here, $\kappa$ is a constant `gain' parameter. The pressure output $\Delta p$ depends on $\theta$ as well as the location of the pressure sensors $\sigma$ and the swimming speed $V$.
Equations~\eqref{eq:eom} can be thus viewed as a model of the neuro-mechanic control or response of the fish to sensory information $\Delta p$.

\section{Results}

We first examine the dependence of the sensory output $\Delta p$ on the  location $\sigma$ of the sensors as the body changes  orientation relative to the oncoming flow and we compare our findings to the experimental results of~\cite{Ristroph2015}. 
One of the advantages of the modeling approach employed here is that it is amenable  to analytic expressions for the  pressure (expression not shown) and the pressure difference $\Delta p$. One gets, upon substituting~\eqref{eq:uniform} into~\eqref{eq:vel} and using the resulting expression in~\eqref{eq:dp0}, that
\begin{equation}
\Delta p = \frac{2R^2\sin(2\sigma)}{R^4-2R^2\cos(2\sigma)+1}  \Bigl[V(R^2-1)\sin(\theta)+R^2\sin(2\theta)\Bigr].
\label{eq:dp}
\end{equation}
Note that the dependence of the function $\Delta p$ on $\sigma$ is separable from its dependence  on  $V$ and  $\theta$. This property has important consequences on optimal sensor placement as discussed below. 


Figure~\ref{fig:dp}(a) shows the pressure distribution for the right and left sensors as a function of $\sigma$ when the body is oriented at $\theta=10\degree$.  The flow-facing (right) side experiences higher pressure than the leeward (left) side, consistent with~\cite[figure 3]{Ristroph2015}. 
 Figure~\ref{fig:dp}(b) depicts $\Delta p$ as a function of $\sigma$ for three different body orientations $\theta = 5, 10$ and $20\degree$. The pressure difference increases with increasing $\theta$ but the dependence on $\sigma$ follows a similar trend for all $\theta$, which is to be expected from the separability property of $\Delta p$ in~\eqref{eq:dp}.  The increase in the sensory signal $\Delta p$ as $\theta$ increases can be interpreted by the swimmer as an indication of the degree of misalignment with the oncoming flow, as noted in~\cite{Venturelli2012,Ristroph2015}. 
The qualitative agreement of the results in figure~\ref{fig:dp}(a,b) with the experimental findings in figure 3(a,b) of~\cite{Ristroph2015} is remarkable (the figures of Ristroph {\em et al.}  are not reproduced here). To further explore the accord between our model and the experimental results, we plot in figure~\ref{fig:dp}(c) the {\em pressure stimulation} parameter, defined  in Ristroph {\em et al.}  as the ratio of the pressure difference to orientation angle $\Delta p /\theta$. The curves in figure~\ref{fig:dp}(c) collapse into one curve for small $\theta$ as in~\cite[figure 3]{Ristroph2015}. Indeed, for small $\theta$,  $\Delta p/\theta$ is independent of $\theta$ as can be readily verified from~\eqref{eq:dp} and, thus, the sensory output $\Delta p$ scales linearly with $\theta$. The collapsed curve shows a remarkable resemblance to the average canal density itself, shown in figure~\ref{fig:lateral-lines-schematic}(b) and overlaid on the stimulation curves in figure~\ref{fig:dp}(d) for direct comparison. This correspondence confirms the findings of~\cite{Ristroph2015} that the lateral line canals are concentrated at locations of strong hydrodynamic signals $\Delta p$. %
%
%
%

We now go beyond the results of~\cite{Ristroph2015} to explore the effect of the swimming speed $V$ on $\Delta p$, as shown in Figure~\ref{fig:dp}(d) 
 for four different speeds $V = 5, 10, 20$ and $40$. Clearly,  $\Delta p$ follows a similar trend for all $V$ such that  the curves $\Delta p / V$ (for non-zero $V$) collapse onto one curve for large $V$.  The fact that for large $V$,  $\Delta p/V$ is independent of $V$ follows immediately from~\eqref{eq:dp}. It implies that for swimming speeds $V$ much larger than the speed of the background flow (here normalized to 1), the sensory output $\Delta p$ scales linearly with $V$. Further, the value of $V$ does not alter the sensors location $\sigma$ for which $\Delta p$ is maximum.
To  compute  analytically the optimal sensor location $\sigma_{\rm opt}$  for which $\Delta p$ takes its maximum value, we set $\partial  \Delta p / \partial \sigma = 0$ to find that
\begin{equation}
\sigma_{\rm opt} =  \frac{1}{2}\cos^{-1}\left( \frac{2R^2}{R^4+1} \right).
\label{eq:opt} 
\end{equation}
 It follows immediately from the separability property in~\eqref{eq:dp0} that $\sigma_{\rm opt}$ is independent of $V$ and $\theta$ and depends only on $R = \sqrt{(a+b)/(a-b)}$.

\begin{figure}
\centering
\includegraphics[width=6in]{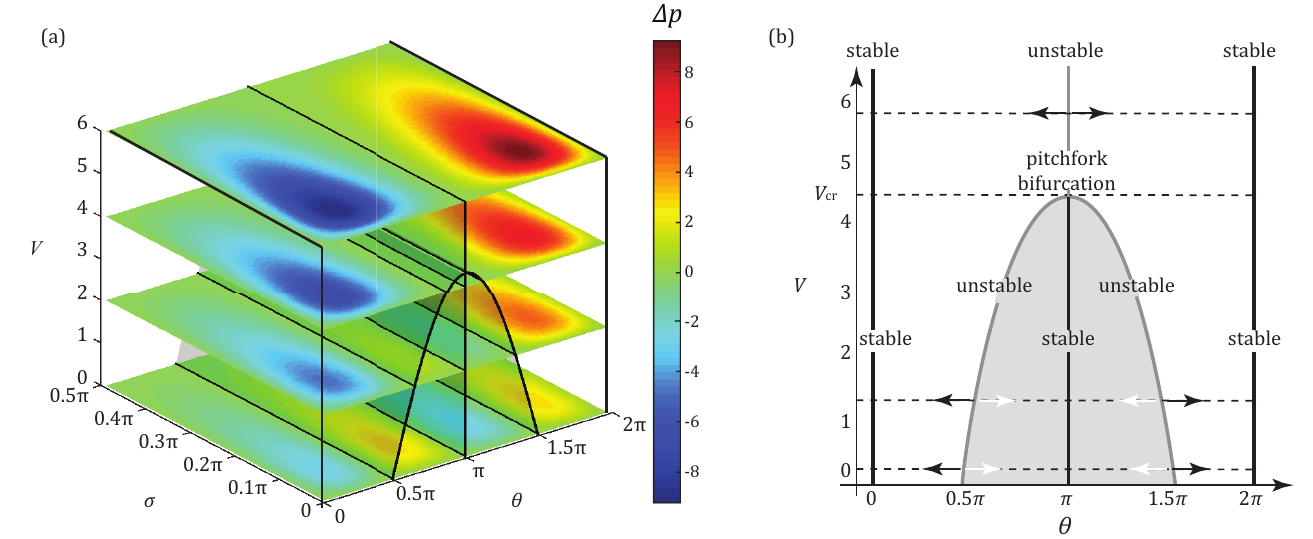}
\caption{(a) Pressure difference $\Delta p$ as a function of body orientation $\theta$, swimming speed $V$ and sensor location $\sigma$. Black contour lines show $\Delta p = 0$. (b) $\Delta p = 0$ are fixed points of the governing equations for which the body does not change orientation. The stability of these fixed directions depend on the swimming speed $V$. The ellipse aspect ratio is set to $b/a = 0.3$, but the results hold qualitatively for all $0<b/a <1$.
 }
\label{fig:dp2}
\end{figure}

In Figure~\ref{fig:dp2}(a) is a depiction of the pressure difference $\Delta p$ as a function of body orientation $\theta$, swimming speed $V$ and sensor location $\sigma$. This pictorial depiction confirms that  the optimal sensor location $\sigma_{\rm opt}$ for which $\Delta p$ is maximum/minimum is independent of $V$ and $\theta$.  Meanwhile, there exist orientations for which the sensory output is identically zero irrespective of sensor placement $\sigma$. These \textit{blind spots}, orientations  for which the sensory output is zero, are highlighted in black in Figure~\ref{fig:dp2}(a) and calcuated analytically by setting $\Delta p =0$ in~\eqref{eq:dp} and solving for $\theta$, which yields
\begin{equation}
\theta^\ast =  0, \qquad \theta^\ast = \cos^{-1}\left(-V \frac{R^2 - 1}{2R^2}\right), \quad \theta^\ast = \pi, \qquad  (\textrm{mod} \ 2 \pi).
\label{eq:blind}
\end{equation}
That is, $\Delta p$ is zero for $\theta^\ast = 0$ and $\pi$ when the body is aligned opposite to or in the same direction as the oncoming uniform flow. A third blind spot exists only when $-1 \leq -V (R^2-1)/2R^2\leq 1$, that is, when the swimming speed  $V$ is below a critical value $V_{\rm cr} =  {2R^2}/({R^2 - 1})$.  

We examine the behavior  when the body changes its orientation in response to the local pressure difference $\Delta p$ according to~\eqref{eq:eom}. Note that  $\Delta p$ and, consequently,  the right-hand side of~\eqref{eq:eom}, depends explicitly on $\theta$, but not on the position $(x_c,y_c)$ of the body. This fact is due to the translational symmetry of motion in uniform flows. One can thus `reduce' equations~\eqref{eq:eom} from three equations for $(x_c,y_c,\theta)$ to one equation in terms of $\theta$ only in the sense that one can solve the orientation equation $\dot{\theta} = \kappa \Delta p$ separately. Once the orientation is known, the trajectory in the $(x_c,y_c)$ plane can be readily reconstructed. 

\begin{figure}[b]
\centering
\includegraphics[width=6in]{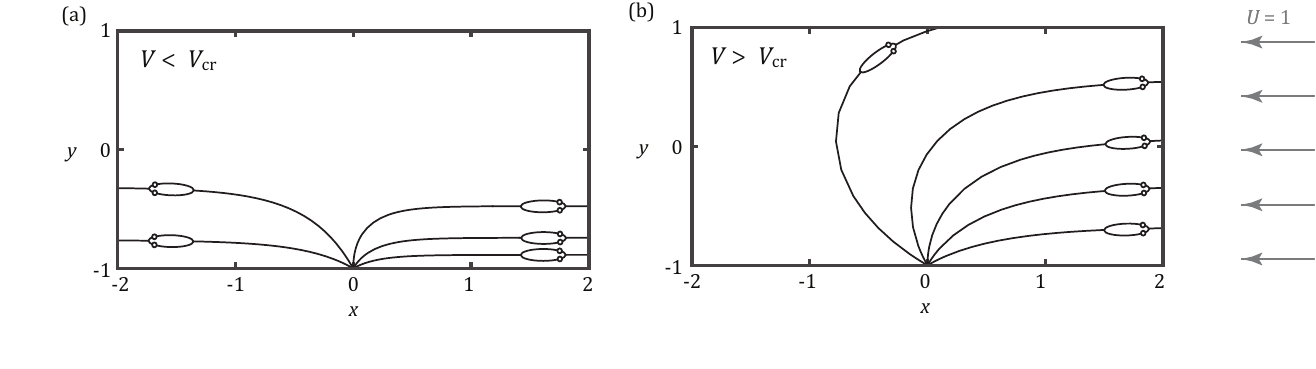}
\caption{Trajectories of swimmer in uniform flow for (a) $V = 1.44 $ and (b) $V = 11.5 $. Initial orientations are $\theta(0) = {\pi/6}, {\pi/3}, {\pi/2}, {2\pi/3}, {5\pi/6}$. Parameter values are $b/a = 0.3$, $V_\textrm{cr} = 4.33$, and $\kappa = 1$.}
\label{fig:uni_traj}
\end{figure}

 
Orientations for which $\Delta p = 0$ are fixed points or, more precisely, relative equilibria of~\eqref{eq:eom}. Along these directions, the body moves with constant speed $V$ without changing orientation.  These directions are linearly stable when $\partial \Delta p / \partial \theta < 0$ and unstable otherwise. A summary of the three families of fixed points and their stability type is given in figure \ref{fig:dp2}(b). For swimmers moving at speeds $V<V_{\rm cr}$, both  $\theta^\ast =0$ and $\theta^\ast = \pi$ are  stable, whereas $\theta^\ast = \cos^{-1}(V/V_{\rm cr})$ is unstable. At $V = V_{\rm cr}$, the system undergoes a \textit{subcritical pitchfork} bifurcation where the two unstable branches of  $\theta^\ast = \cos^{-1}(V/V_{cr})$ collide with the stable branch $\theta^\ast = \pi$, making it unstable. For $V>V_{\rm cr}$, one has only one stable solution $\theta^\ast = 0$ where the body aligns in the opposite direction to the oncoming flow. In other words, for low swimming speeds $V$ relative to the normalized background flow, the swimmer orients either with or against the flow. As the swimming speed increases, the swimmer orients only in the opposite direction to the flow. Figure~\ref{fig:uni_traj} depicts representative trajectories for both $V <V_{\rm cr}$ and $V> V_{\rm cr}$. These results are consistent with observations of fish rheotaxis where 
 fish are seen to respond to water motion either by fleeing high-flow regions or by aligning their bodies in the direction opposite to the oncoming flow for upsteam migration and position holding~\cite{Arnold1974, Montgomery1997}.
 
\begin{figure}
\centering
\includegraphics[width=6in]{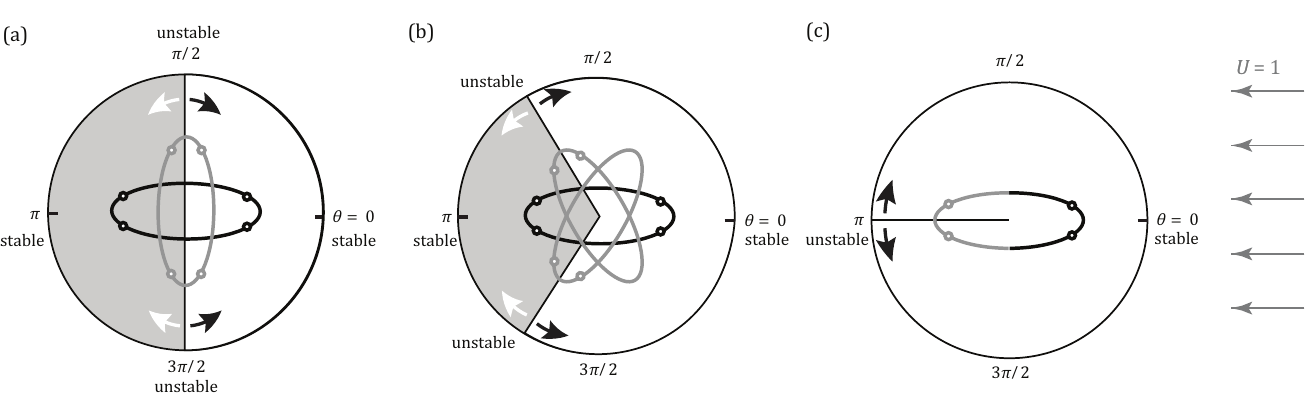}
\caption{Basins of attraction: (a) For $V = 0$, the set of all initial orientations are equally split into a white subset  that tends to align  in the direction opposite to the incoming flow, and a grey subset  that tends to align in the same direction as the incoming flow. (b) For $0<V<V_{\rm cr}$, the grey region shrinks until at $V=V_{\rm cr}$, a pitchfork bifurcation occurs making $\theta^\ast = \pi$ unstable. (c) For   $V>V_{\rm cr}$, all trajectories, except those parallel to $\pi$, tend to align in the direction opposite to the flow.}
\label{fig:BasinsofAttraction}
\end{figure}

As time increases, the linearly stable direction $\theta^\ast = 0$ is an {\em attractor} for all $V$ and $\theta^\ast = \pi$ is an attractor for $V < V_{\rm cr}$. For each stable direction $\theta^\ast$, the \emph{basin of attraction} is defined as the set of all initial values of $\theta$ for which the body eventually aligns with $\theta^\ast$. 
Figure~\ref{fig:BasinsofAttraction}  depicts the basins of attraction for three representative cases corresponding to (a)  zero swimming speed $V =0$, (b) swimming at subcritical speeds $V<V_{\rm cr}$, and (c) swimming at supercritical speeds $V>V_{\rm cr}$. In (a) and (b), the set of all orientations is divided into two regions: the white region defines the basin of attraction for the stable orientation $\theta^\ast = 0$, where the body aligns in the opposite direction to the flow,  and the grey region defines the basin of attraction for $\theta^\ast =\pi$ for which the body aligns with the flow. For $V>V_{\rm cr}$, the latter orientation becomes unstable and all initial conditions tend to align in the direction  opposite to the oncoming flow.

\section{Discussion}

The lateral line sensory system, which contains canal receptors that measure hydrodynamic pressure, is thought to be responsible for the ability of fish to sense minute water motions in their vicinity, and, thus, to play
an important role in the fish response to ambient flows such as fish rheotaxis.  Recent findings based on fish anatomical measurements and flow experiments suggest that the arrangement of the lateral line system is correlated with locations along the fish body that experience strong variations in the hydrodynamic signal~\cite{Ristroph2015}. 

In this communication, we presented a mathematical model consisting of an elliptic body equipped with laterally-arranged pressure sensors that emulate the fish lateral line sensory system. We studied the sensory output (the difference in hydrodynamic pressure) in oncoming uniform flows as a function of the sensors location, swimming speed and body orientation. We found that the optimal placement of sensors that maximizes the sensory output is independent of the swimming speed and body orientation and depends only on the body geometry. Our results are consistent with the experimental findings of~\cite{Ristroph2015} and support a physical and hydrodynamic explanation of the highly conserved architecture of the lateral line sensory system across many fish species -- the sensory system extends over much of the body but is concentrated near the head. Most fish species have a body plan characterized by high curvature at the head. This body plan leads to a pressure profile that is greatest at the head, making it particularly sensitive to variations in the hydrodynamic signal, such as those due to changes in body orientation or swimming speed.

For small deviations from the oncoming flow directions, the sensory output varies linearly with the body orientation. This linear dependence holds well up to $\theta = 20\degree$. The  pressure output  also changes linearly with the swimming speed $V$ when $V$ is much larger than the background flow $U=1$, namely, when $V/U > 40$. Such linear encoding of changes in sensory signal as the fish swimming parameters change could greatly simplify the neural processing involved in signal decoding. However, this linear dependence does not hold for  finite body orientation and relatively moderate swimming speeds, implying that more complex computations are needed to decipher the sensory information in these regimes. 

We proposed that the fishlike body responds to these sensory cues by adjusting its orientation proportionally to the perceived difference in pressure. This modeling approach can be viewed as an abstraction, or `meta-model', of the coupling between the sensory output and the neural and muscular drive underlying fish behavior. Our working hypothesis is that these intertwined sensory, neural and actuation mechanisms produce an emergent rule or law  that underlies fish rheotaxis.  The behavior obtained from our proposed model is remarkably similar to the fish rheotactic behavior. At relatively low swimming speeds, fish go with the flow if they are initially oriented in a way that they cannot hold out against and turn into the oncoming flows. Otherwise, they align into the oncoming flows.



This modeling framework couples the sensory and response mechanisms underlying fish rheotaxis,~albeit in the context of a potential flow model. It establishes the response law to sensory signals, regardless of the physiological  mechanisms that bring this about. Here, a few comments on the model limitations are in order:
\vspace{-1ex}
\begin{itemize}
\item[](i) We considered a planar model. 
But 
an ellipsoidal body in uniform potential flows leads to similar results on optimal sensor location, equilibrium orientations, and stability.
In uniform flows, the  flow topology and pressure distribution around  the centerline of an ellipsoidal body are similar to those around an ellipse -- this is the reason why our results agree with the experimental findings of~\cite{Ristroph2015}. 
\item[](ii) We ignored the effects of viscosity and separation.   In the experiments of~\cite{Ristroph2015} where the fish model is placed in uniform flows, separation  occurs at the trailing edge of the fish -- thus it does not affect the optimal location of sensors. Viscosity and separation  will play prominent roles in non-uniform flows and complex fish maneuvers. Under these conditions, fish need multiple sensory outputs from their distributed lateral line to decipher the location of flow separation. 
\item[](iii) The  control law in~\eqref{eq:eom} is linear. This is in part due to the lack of experimental evidence to support a particular choice of linear versus nonlinear control law.  More importantly, a linear control law simplifies the neural processing involved in responding to a given sensory output. Therefore, it is reasonable to conjecture that a linear law is preferable. We showed that  the linear model leads the proper rheoactic behavior. More complex fish behavior might require more complex control laws.
\item[](iv) The model in~\eqref{eq:eom} is kinematic. That is to say, it does not explicitly account for forces and moments, although they can be calculated posteriori for each trajectory using balance of momenta. Kinematic models are common in studying emergent behavior such as in fish schooling~\cite{Couzin2002} where details of the coupling among the sensory feedback, neural control, and muscular actuation is not fully understood. Our goal here was to test whether the rule `turn in response to sensory output'  reproduces fishlike rheotaxis. We will  link this rule-based model to the mechanics of motion in ambient flows in future studies.
\end{itemize}

\vspace{-1ex}
The two main results obtained from the simplified model --  the optimal sensor placement that maximizes the sensory output and how rheotactic behavior emerges in light of this sensory information --  serve to direct future research. We focused here on single sensory output and rheotactic behavior in uniform flows. But future work will address the interplay of multiple sensory outputs and how such sensory information could be used to avoid collision with nearby objects, detect ambient vorticity, and inform changes in the fish body undulations~\cite{Fernandez2011,Ren2012, Yanase2012}. Future models will also account for three-dimensional and viscous effects~\cite{Hassan1992,Eldredge2008,Gazzola2011}, which we expect to play a prominent role in non-planar fish motions and maneuvers as well as in non-uniform and high-speed flows, where flow separation is paramount. 
These models could also serve as a design tool for developing optimal sensory layouts and deployment strategies of underwater vehicles \cite{Fernandez2011, Triantafyllou2016, Venturelli2012,Yang2006}. The development of autonomous vehicles that detect hydrodynamic cues is important for their deployment in unknown underwater environments and is relevant for many environmental, military, and industrial applications including ocean exploration, navigation, surveillance and monitoring.



%



\paragraph{Acknowledgement.}
This work is partially supported by the ONR grant 14-001. Brendan Colvert acknowledges the support of the 
Department of Defense (DoD)
through a National Defense Science and Engineering Graduate (NDSEG) Fellowship. 

\bibliographystyle{plain}
\bibliography{ColvertKanso2016_jfm}

\end{document}